\documentclass[10pt,aps,prapplied,twocolumn,longbibliography,nobibnotes,showkeys]{revtex4-2}

\usepackage{graphicx}
\usepackage{dcolumn}
\usepackage{bm}
\usepackage[version=3]{mhchem}
\setcitestyle{super,open={},close={}}

\begin{document}
\title{Hydrogen defects as probes of band alignment in metal-organic frameworks}
\author{Khang Hoang}
\email{khang.hoang@ndsu.edu}
\affiliation{Center for Computationally Assisted Science and Technology \& Department of Physics, North Dakota State University, Fargo, North Dakota 58108, United States}

\date{\today}

\begin{abstract}

Band alignment, namely the prediction of band-edge positions of semiconductors and insulators in aqueous solutions, is an important problem in physics and chemistry. Such a prediction is especially challenging for structurally and chemically complex, multi-component materials. Here we present an approach to align band structure of metal-organic frameworks (MOFs) on an absolute energy scale which can be used for direct comparison with experiments. Hydrogen defects are used as probes into the chemical bonding of the hybrid inorganic-organic materials. An effective hydrogen defect level, defined as the average of the charge-state transition levels of the defects at the secondary building unit and at the linker, is identified as a charge neutrality level to align band structures. This level captures subtle chemical details at both the building blocks and provides results that are in agreement with experiments in a wide range of different MOFs. We also compare with results obtained from using other approaches involving surface calculations and average pore-center electrostatic potentials.  

\end{abstract}


\maketitle


\section{Introduction}\label{sec;intro}

Metal-organic frameworks (MOFs) are covalently bonded hybrid materials composed of inorganic secondary building units (SBUs, which are metal atoms or clusters) and organic linkers. Due to their high and tunable porosity, large surface area, structural flexibility, physical and chemical stability, and variable chemical functionality, MOFs have been of interest for gas storage, gas separation, and catalysis, among many others.\cite{Furukawa2013Science,Li2018AM,Chueh2019JMCA,Navalon2023CR} In applications such as photocatalysis, photovoltaics, and electrochemistry, the ability to predict band-edge positions of MOFs on an absolute energy scale is essential to understanding the underlying mechanisms and to materials design. In catalysts or sensors based on MOF/oxide composites,\cite{Chen2025ChemEurope} for example, proper band alignment that ensures efficient charge separation and transfer is critical to their performance. Yet, as discussed below, reliable prediction of band alignment in solids is inherently challenging; it is more so in MOFs.

From the perspective of band theory, MOFs can be treated as conventional semiconductors or insulators.\cite{Mancuso2020CR,Kolobov2021AC} One can predict the valence-band maximum (VBM) and conduction-band minimum (CBM) with respect to the vacuum level by applying the slab approach which involves both bulk and surface (slab) calculations.\cite{Franciosi1996SSR,Moses2010APL} Band edges predicted from such an approach can, however, be highly sensitive to details of surface termination and composition. Computational treatment of MOF surfaces using first-principles calculations are also challenging due to the large unit cell sizes, making the exploration of multiple surfaces and surface terminations impractical. On the other hand, MOF surfaces in real samples, on which electrochemical measurements of band edges must be performed, are often ill-defined. Besides, one may wonder if the consideration of external MOF surfaces is at all justified when internal MOF surfaces are vast and even more relevant in measurements and in applications. In that context, Butler et al.'s approach\cite{Butler2014JACS} offers a simple solution according to which the vacuum level can be determined from the average electrostatic potential at the center of a MOF pore. This approach, however, works only for porous MOFs with large pore sizes. As will be shown later, both these approaches do not perform well when results for MOFs are directly compared with experiments. 

Here we develop an approach inspired by and based on the universal alignment of hydrogen levels in semiconductors, insulators, and solutions discovered by Van de Walle and Neugebauer\cite{VdW2003Nature} and related to the concept of charge neutrality level (CNL).\cite{Tejedor1977JPC,Flores1979JPC,Tersoff1984PRL,Cardona1987PRB,WalukiewiczPRB1988,VdW2003Nature,King2009PRB,Varley2024JAP} In elemental and simple compound semiconductors and insulators, CNL is often defined as an effective energy level at which the bulk electronic states are equally valence-band-like (or donor-like) and conduction-band-like (acceptor-like); it has been argued to be an intrinsic property of the materials. Energies that can be identified with CNL such as the charge-state transition level of interstitial hydrogen defects have been shown to lie at a common reference level from which band edges in different materials can be aligned.\cite{VdW2003Nature,Peacock2003APL,King2009PRB,Varley2017JPCL,Swallow2019APLM} The hydrogen interstitial $(+/-)$ transition level, which probes the anion- and cation-derived dangling bond states in a compound semiconductor (formed by the strong interaction between the hydrogen and the host material), was found to be approximately at 4.5 eV below the vacuum energy level [which can be identified with the H$^+$(aq.)/H$_2$(g) electrode in water, 4.44 eV below the vacuum level].\cite{VdW2003Nature} In addition to the hydrogen defect levels, energies directly associated with dangling bonds in materials can also serve as a CNL.\cite{Lefebvre1987PRB,WalukiewiczPRB1988,VdW2003Nature,Varley2024JAP} In our approach, we make use of the charge-state transition levels of hydrogen interstitials at the SBU and at the linker in MOFs. An effective defect level, which captures the chemical bonding at both the inorganic and organic building blocks, is identified as a CNL which is then used to align the calculated band edges.   

We demonstrate our approach on two series of MOFs; one includes PCN-222(2H) and its metal-substituted analogs PCN-222($M$) with $M$ = Mn, Fe, Ni, Cu, Zn, or Pt, and the other includes MOF-5, MIL-125, UiO-66, and ZIF-8. PCN-222(2H), also known as MOF-545 or MMPF-6, has the space group $P_6/mmm$ and is composed of a Zr$_6$O$_8$-based cluster as SBU and tetrakis(4-carboxyphenyl)-porphyrin (H$_2$-TCPP) as linker.\cite{Morris2012IC,Feng2012AC,Chen2012IC} PCN-222($M$) has the same crystal structure but with the two H atoms in the inner ring of the TCPP linker being replaced with $M$. The difference between the compounds in the PCN-222 series is thus just at the organic linker. In the second series, MOF-5 ($Fm\bar{3}m$) with the formula Zn$_4$O(BDC)$_3$ has the Zn$_4$O cluster as SBU and 1,4-benzenedicarboxylate (BDC) as linker.\cite{Li1999Nature} MIL-125 ($I_4/mmm$) has the formula Ti$_8$($\mu_2$-O)$_8$($\mu_2$-OH)$_4$(BDC)$_6$,\cite{DanHardi2009JACS} and UiO-66 ($Fm\bar{3}m$) is Zr$_6$($\mu_3$-O)$_4$($\mu_3$-OH)$_4$(BDC)$_6$,\cite{Cavka2008JACS} where $\mu_2$-O or $\mu_3$-O is a structure in which the oxygen atom is bridging two or three metal atoms, respectively. Finally, ZIF-8 ($I\bar{4}3m$) is a zeolitic imidazolate framework, C$_8$H$_{10}$N$_4$Zn, with the Zn$^{2+}$ ion regarded as SBU and 2-methylimidazolate (2-mIM) as linker.\cite{Park2006PNAS}

\section{Methodology}\label{sec;method} 

First-principles calculations are based on density-functional theory (DFT) with the projector augmented wave method\cite{PAW1,PAW2} and a plane-wave basis set, as implemented in \textsc{vasp}.\cite{VASP2} For PCN-222(2H) and PCN-222(Zn), we use the Perdew-Burke-Ernzerhof (PBE) functional,\cite{GGA} whereas for PCN-222($M$) with $M = $ Mn, Fe, Ni, Cu, or Pt, we use the PBE$+$$U$ method~\cite{anisimov1991} with the effective Hubbard $U$ parameter for the transition-metal $d$ orbitals set to 6 eV for $M$ = Fe, Ni, or Pt, 7 eV for Cu, and 10 eV for Mn. For MOF-5, MIL-125, UiO-66, and {ZIF-8}, we use the the Heyd-Scuseria-Ernzerhof (HSE) hybrid functional\cite{heyd:8207} with the lattice constants calculated at the PBE (for MOF-5 and UiO-66), PBE$+$$U$ (MIL-125; $U$ = 6 eV), or HSE (ZIF-8) level; the internal coordinates are all fully relaxed within HSE. The screening length is set to the standard value of 10 {\AA}, whereas the Hartree-Fock mixing parameter ($\alpha$) is adjustable. The density functional and associated parameters are chosen to reasonably reproduce the electronic structure and to match the experimental band gap reported in the literature. In all the calculations, the energy cutoff is set to 500 eV and spin polarization is included. Structural relaxations are performed with the force threshold chosen to be 0.02 eV/{\AA}. Only the $\Gamma$ point is used for the integration over the Brillouin zone. A denser {\rm k}-point mesh is used to obtain high-quality density of states.

We model interstitial hydrogen defects (H$_i$) in MOFs using a supercell approach in which an extra hydrogen atom is included in a periodically repeated finite volume of the host materials. The supercell size is from 240 host-atoms (MIL-125) to 630 host-atoms [PCN-222(2H)] plus the interstitial hydrogen atom. H$_i$ can exist in the neutral, positive, or negative charge state, which is simulated by controlling the total number of valence electrons in the supercell; \textsc{vasp} automatically applies a compensating homogeneous background charge over the entire charged supercell to eliminate Coulomb divergence. In the defect calculations, the lattice parameters are fixed to the optimized bulk values, but all the internal coordinates are relaxed. This is to avoid spurious elastic interactions with defects in neighboring supercells. 

The formation energy of H$_i$ in effective charge state $q$ (where $q = +, 0, -$ with respect to the host lattice) is defined as\cite{walle:3851,Freysoldt2014RMP}     
\begin{align}\label{eq:eform}
E^f({\mathrm{H}_i}^q)=&&E_{\mathrm{tot}}({\mathrm{H}_i}^q)-E_{\mathrm{tot}}({\mathrm{bulk}}) -\mu_{\rm H}^\ast \\
\nonumber &&+ q(E_{\mathrm{v}}+\mu_{e}) + \Delta^q,
\end{align}
where $E_{\mathrm{tot}}(\mathrm{H}_i^{q})$ and $E_{\mathrm{tot}}(\mathrm{bulk})$ are the total energies of the defect-containing and perfect bulk supercells, respectively. $\mu_{\rm H}^\ast = \frac{1}{2}E(\rm {H_2}) + \mu_{\rm H}$ is the hydrogen chemical potential, representing the energy of the reservoir with which a hydrogen atom is being exchanged; $E(\rm {H_2})$ is the total energy of an isolated H$_2$ molecule at 0 K. In the following presentation and unless otherwise noted, we assume that the MOF host is in equilibrium with H$_2$ gas at 1 bar and 400 K which results in $\mu_{\rm H} = -0.23$ eV, based on the tabulated data in Ref.~\citenum{H2gas}. This choice is somewhat arbitrary as we are not interested in the actual value of the formation energy. $\mu_{e}$ is the chemical potential of electrons, i.e., the Fermi level, representing the energy of the electron reservoir, referenced to the VBM in the bulk ($E_{\mathrm{v}}$). $\Delta^q$ is the correction term to align the electrostatic potentials of the bulk and defect supercells (to eliminate the spurious effects caused by the application of the neutralizing background mentioned above) and to account for finite-size effects on the total energies of charged defects, following the approach of Freysoldt et al.\cite{Freysoldt} as implemented in the {\it sxdefectalign} code.\cite{sxdefectalign}

The thermodynamic transition level between charge states $q_1$ and $q_2$ of H$_i$, $\epsilon(q_1/q_2)$, is defined as the Fermi-level position at which H$_i^{q_1}$ and H$_i^{q_2}$ have equal formation energies, i.e.,~\cite{Freysoldt2014RMP}
\begin{equation}\label{eq;tl}
\epsilon(q_1/q_2) = \frac{E^f({\rm H}_i^{q_1}; \mu_e=0)-E^f({\rm H}_i^{q_2}; \mu_e=0)}{q_2 - q_1},
\end{equation}
where $E^f({\rm H}_i^{q}; \mu_e=0)$ is the formation energy of H$_i^q$ when the Fermi level is at the VBM ($\mu_e=0$). This $\epsilon(q_1/q_2)$ level, corresponding to a defect level, would be observed in experiments where the defect in the final charge state $q_2$ fully relaxes to its equilibrium configuration after the transition. Note that $\epsilon(q_1/q_2)$ is independent of the choice of the atomic chemical potential.   

The calculation of $\Delta^q$ in Eq.~(\ref{eq:eform}) requires knowledge of the total (electronic + ionic) static dielectric constant ($\epsilon_0$), which can be obtained from the macroscopic ion-clamped static dielectric tensor.\cite{Gajdos2006PRB} We find $\epsilon_0^{xx} = \epsilon_0^{yy} = \epsilon_0^{zz} = 1.72$ in MOF-5, 3.63 in UiO-66, or 2.90 in ZIF-8, whereas $\epsilon_0^{xx} = \epsilon_0^{yy} = 3.27$ and $\epsilon_0^{zz} = 2.26$ in MIL-125. For the PCN-222 series, due to the exceedingly demanding computational cost of the dielectric tensor calculation, we assume $\epsilon_0^{xx} = \epsilon_0^{yy} = \epsilon_0^{zz} = 3.00$ in the calculation of $\Delta^q$, comparable to the experimental value of $\sim$2.3 measured at 10$^3$ Hz in solvent-free Al-PMOF~\cite{Zhang2021CC}--another MOF which also possesses the TCPP linker. Since in this work we are interested in the effective $(+/-)$ level, which is the average of the local $\epsilon(+/-)$ levels at the SBU and at the linker, any deviation from actual $\epsilon_0$ values for PCN-222 would result only in negligible changes in the calculated $(+/-)$ level. This is because $\Delta^q$ is largely canceled out in the total-energy difference involving H$_i^+$ and H$_i^-$. We find that the correction to the effective $(+/-)$ level is only 20--30 meV in the case of PCN-222, MIL-125, and {ZIF-8} or 50--60 meV in MOF-5 and UiO-66.

Finally, to align the band edges on an absolute energy scale, we assume the universal alignment of hydrogen defect levels.\cite{VdW2003Nature} For comparison, band alignment is also determined by referencing to the vacuum which involves both bulk and surface calculations\cite{vdW1986PRB,VdW1989PRB,VandeWalle2003JVSTA} or by referencing the DFT-calculated band edges to the spherical average of the total electrostatic potential at the center of the pore of the MOFs using the MacroDensity code.\cite{Butler2014JACS,Harnett2021JCP} Slab calculations for the (001) surface of PCN-222(2H) and PCN-222(Cu) are carried out using supercells of 2358 and 2334 atoms, respectively, and a vacuum layer of about 32 {\AA}; see the slab models in Fig.~S1 in Supplementary Information. An energy cutoff of 300 eV and only the $\Gamma$ point are used in these slab calculations.

\section{Results and Discussion}\label{sec;results}

In the following, we present results for bulk properties of the MOFs obtained in first-principles calculations, with a focus on their atomic and electronic structure, which serve as a basis for the understanding of hydrogen defects and the determination of electronic band alignment. We then discuss structural, electronic, and energetic properties of hydrogen interstitials in the MOF hosts. Finally, we discuss band alignment predicted using our approach based on hydrogen defect levels and compare with experiments\cite{Semerci2023CT,Hariri2021AOC,Xu2015JACS,Jin2020NJC,Li2023P,Yuan2023CSA,Zhang2022JC,Duan2020APCS,Huang2024SCM,Aleksandrzak2020RSCA,Chen2021RKMC,Yao2022JSSC,Chen2020IECR,Jin2024RSCAdv,Dai2025MD,Hou2022CST,Sun2024IC,Mu2018DT,Gayathri2022CSA,Xu2018ACSC,Li2022CST} and with that obtained using existing approaches.

\subsection{Bulk atomic and electronic structure}\label{sec;results;bulk}

\begin{figure}
\vspace{0.2cm}
\includegraphics[width=\linewidth]{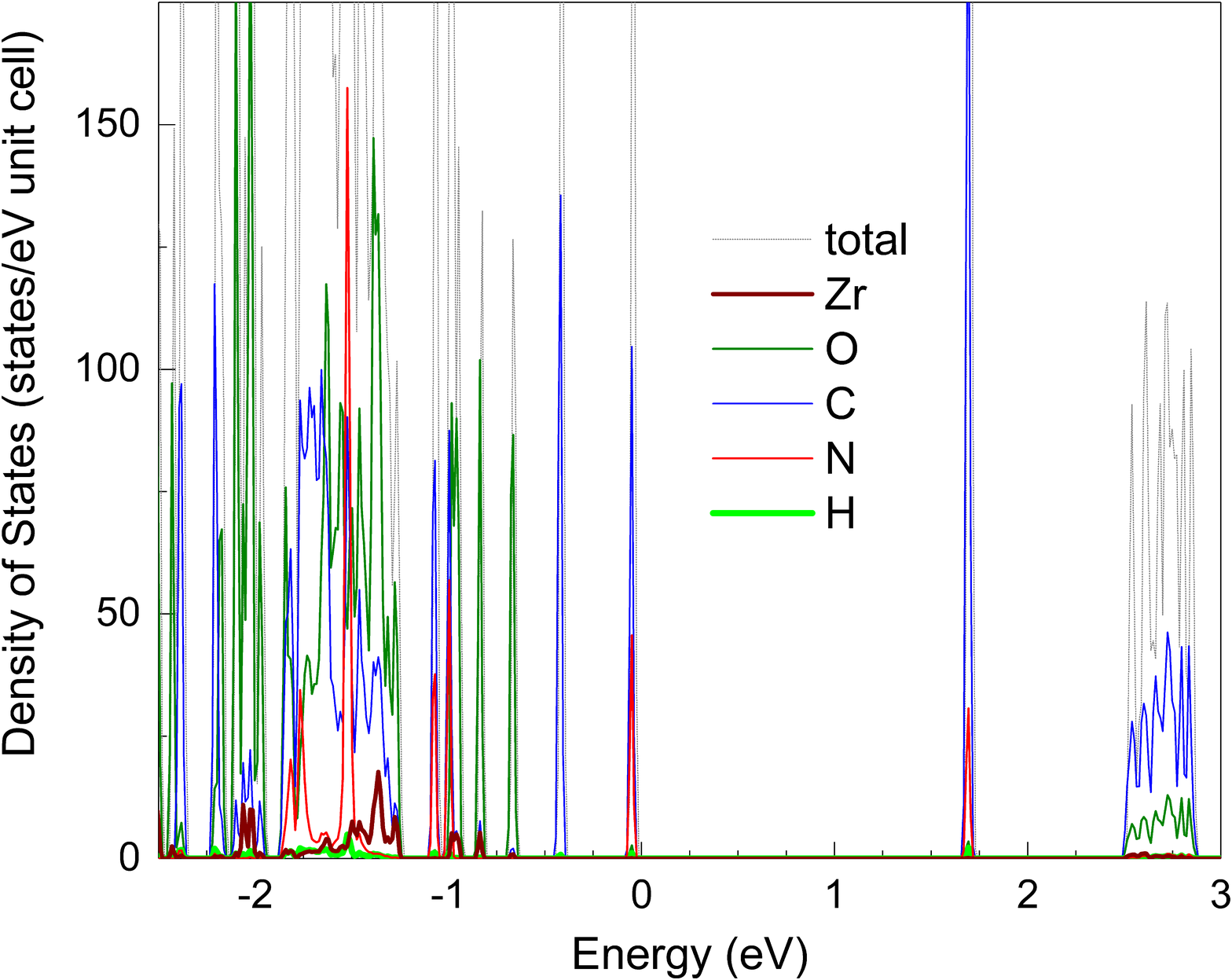}
\caption{Total and atom-decomposed electronic densities of states in PCN-222(2H). The zero of energy is set to the highest occupied state.} 
\label{fig;dos} 
\end{figure}

Different proton topologies have been proposed for the Zr$_6$O$_8$-based SBU in PCN-222(2H). These include [Zr$_6$($\mu_3$-O)$_8$(H$_2$O)$_8$]$^{8+}$,\cite{Morris2012IC} [Zr$_6$($\mu_3$-OH)$_8$(OH)$_8$]$^{8+}$,\cite{Feng2012AC} and [Zr$_6$($\mu_3$-O)$_4$($\mu_3$-OH)$_4$(OH)$_4$(H$_2$O)$_4$]$^{8+}$.\cite{Feng2019ACSC} Planas {\it et al.}\cite{Planas2014JPCL} originally proposed the mixed hydroxyl/water topology for Zr$_6$O$_8$-based SBU in NU-1000. We find that PCN-222(2H) with eight terminal water molecules has the lowest energy; PCN-222(2H) with eight bridging and terminal hydroxyl groups or with the mixed hydroxyl/water topology is higher in energy by 1.08 eV or 0.37 eV per SBU, respectively. The proton topology [Zr$_6$($\mu_3$-O)$_8$(H$_2$O)$_8$]$^{8+}$ is thus adopted for bulk PCN-222 compounds in our calculations.

Tables~S1 and S2 summarize the structural and electronic properties of the MOFs. In PCN-222(2H), there are two H atoms in the inner ring of the TCPP linker; Fig.~S2(a). In PCN-222($M$), the two H atoms are substituted by a metal ($M$); see Fig.~S2(b) for $M$ = Ni. $M$ is four-fold coordinated with N in a slightly distorted square planar and most stable as high-spin Mn$^{2+}$, high-spin Fe$^{2+}$, low-spin Ni$^{2+}$, Cu$^{2+}$, Zn$^{2+}$, or low-spin Pt$^{2+}$. In each PCN-222($M$), there are two $M$$-$N bonds with almost equal bond lengths; see Table~S1. The $M$$-$N distance follows the trend of the Shannon ionic radius of the metal ions.\cite{Shannon1976} For all the MOFs, the calculated lattice constants are within 3\% of the experimental values.

\begin{figure*}
\centering
\includegraphics[width=0.95\linewidth]{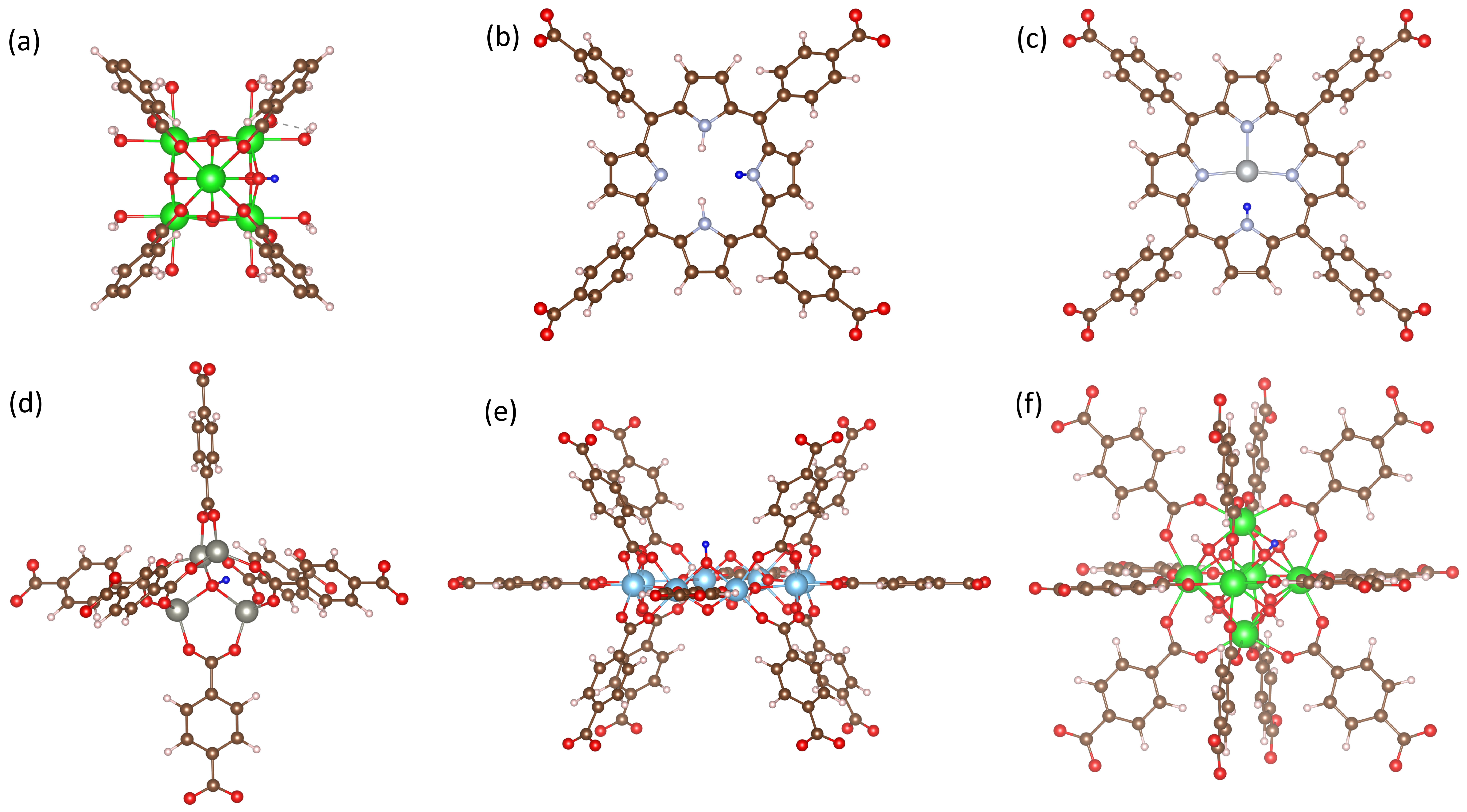}
\caption{Local structure of (a) H$_i^+$ at the SBU in PCN-222(2H); H$_i^-$ at the linker in (b) PCN-222(2H) and (c) PCN-222(Ni); H$_i^+$ at the SBU in (d) MOF-5, (e) MIL-125, and (f) UiO-66, visualized using \textsc{vesta}.\cite{VESTA}. Color code: red = O, brown = C, pink = H, light blue = N, blue = Ti, green = Zr, and gray = Ni. The hydrogen defect is represented by a small blue sphere.} 
\label{fig;hi;struct} 
\end{figure*}

Figure~\ref{fig;dos} shows the density of states of PCN-222(2H); the results for PCN-222($M$) are reported in Fig.~S3. In PCN-222, both the VBM and CBM consist predominantly of the $2p$ states of the C and N atoms residing in the central ring of the TCPP linker; see also the charge density isosurface in Fig.~S2. The central ring would thus play an important role in processes that involve charge transfer. Only in the case of PCN-222(Fe), there is a significant contribution from the Fe $3d$ states to the VBM. For MOF-5, MIL-125, and UiO-66, see Fig.~S4, which have the same BDC linker, the VBM is predominantly the C $2p$ states (and, in the case of UiO-66, a small contribution from the O $2p$ states); the CBM is predominantly the C $2p$ and O $2p$ states (MOF-5), the Ti $3d$ states and some contributions from the C $2p$ and O $2p$ states (MIL-125), or the C $2p$ and O $2p$ states and some contributions from the Zr $4d$ states (UiO-66); Figs.~S5 and S6. In ZIF-8, the VBM and CBM are predominantly the C $2p$ states at the linker; Fig.~S6. 

The band gap is from 1.60 eV for PCN-222(Fe) to 1.98 eV for PCN-222(Pt), 3.84 eV for MOF-5, 3.55 eV for MIL-125, 3.96 eV for UiO-66, and 5.07 eV for ZIF-8; see Tables~S1 and S2. The calculated values are in good agreement with experiments.\cite{Semerci2023CT,Hariri2021AOC,Xu2015JACS,Jin2020NJC,Li2023P,Duan2020APCS,Yuan2023CSA,Zhang2022JC,Huang2024SCM,Aleksandrzak2020RSCA,Chen2021RKMC,Yao2022JSSC,Chen2020IECR,Jin2024RSCAdv,Dai2025MD,Hou2022CST,Sun2024IC,Mu2018DT,Gayathri2022CSA,Xu2018ACSC,Li2022CST,Jing2014RSCAdv} We find that the calculated band gap of PCN-222 is not sensitive to the Hubbard $U$ value used in the PBE$+$$U$ calculations as long as it is in a reasonable range for the transition metals (about 5--7 eV), except in the case of PCN-222(Mn) where a larger $U$ value is needed to have a better match with the experimental band gap. The HSE functional with the standard $\alpha$ value (0.25) is found to overestimate the band gap of PCN-222. For example, HSE ($\alpha=0.25$) calculations give a band gap of 2.25 eV, 2.47 eV, or 2.35 eV for PCN-222(2H), PCN-222(Ni), or PCN-222(Zn), respectively, much larger than that reported in experiments. We use smaller $\alpha$ values in the case of MOF-5 ($\alpha=0.05$), MIL-125 ($\alpha=0.22$), UiO-66 ($\alpha=0.18$), and ZIF-8 ($\alpha=0.20$). 

We note that the band edges as shown, e.g., in Fig.~\ref{fig;dos}, are not yet referenced to any absolute energy level. Also, in the context of this work, it is helpful to emphasize that MOFs are hybrid materials composed of inorganic SBUs and organic linkers, and their atomic and electronic structure can generally be regarded as a superposition of the SBU part and the linker part with the latter having a smaller band gap (i.e., the band edges are predominantly composed of electronic states from atoms at the linker).

\subsection{Interstitial hydrogen defects}\label{sec;results;hi}

We model a hydrogen interstitial (H$_i$) by adding a hydrogen atom to an interstitial site in the bulk supercell. In each MOF, H$_i$ configurations at inequivalent interstitial sites in all possible charge states ($+$, $0$, $-$) are investigated. In the following, we present only the lowest-energy configurations at the SBU and at the linker. 

At the SBU, H$_i$ is bonded to an $\mu_3$-O atom [in the case of PCN-222(2H), PCN-222($M$), and UiO-66; see Figs.~\ref{fig;hi;struct}(a) and \ref{fig;hi;struct}(f)] with the O--H distance of 0.97 {\AA} or to an $\mu_2$-O atom [in the case of MIL-125; Fig.~\ref{fig;hi;struct}(e)] with the O--H distance of 0.96 {\AA}. The hydrogen defect significantly disturbs the metal--oxygen bonding. For example, H$_i^+$ (i.e., the addition of an H$^+$ ion) at the SBU of PCN-222(2H) changes the Zr--O bond lengths at the O site it is attached to from 2.07 {\AA} and 2.13 {\AA} in the perfect bulk to 2.24 {\AA} and 2.43 {\AA}, respectively. H$_i^0$ (H$_i^-$) turns one (two) Ti$^{4+}$ ions in the Ti-based SBU of MIL-125 into Ti$^{3+}$. H$_i$ breaks one of the Zn--O bonds in MOF-5 and forms bonding with the oxygen originally at the center of the Zn$_4$O cluster which now becomes $\mu_3$-like [Fig.~\ref{fig;hi;struct}(d)]; the O--H distance is 0.98 {\AA}. In ZIF-8, H$_i$ breaks one of the Zn--N bonds and forms bonding with the nitrogen (H$_i^+$; N--H: 1.01 {\AA}) or with the zinc (H$_i^-$; N--H: 1.57 {\AA}); Figs.~S7(a) and S7(b). Since the SBU of ZIF-8 contains only the Zn$^{2+}$ ion, H$_i$ at the SBU is thus nominally at the boundary between the SBU and the linker.

At the linker, H$_i$ is bonded to an N atom in the inner ring of the TCPP linker in the case of PCN-222(2H) and PCN-222($M$); see Figs.~\ref{fig;hi;struct}(b) and \ref{fig;hi;struct}(c). The N--H$_i$ bond is out of the linker plane and the angle is different for different PCN-222 compounds. H$_i^0$ (H$_i^-$) even changes the oxidation state of the metal ion at certain $M$-substituted TCPP linkers (specifically, $M$ = Ni or Cu). In the case of MOF-5, MIL-125, UiO-66, and ZIF-8, H$_i$ joins one of the CH sites and forms a CH$_2$ unit that is in perpendicular to the benzene (or imidazole) ring, see Figs.~S7(d) and S7(c), and disturbs the chemical bonding at the BDC (2-mIM) linker.

\begin{figure}
\includegraphics[width=0.98\linewidth]{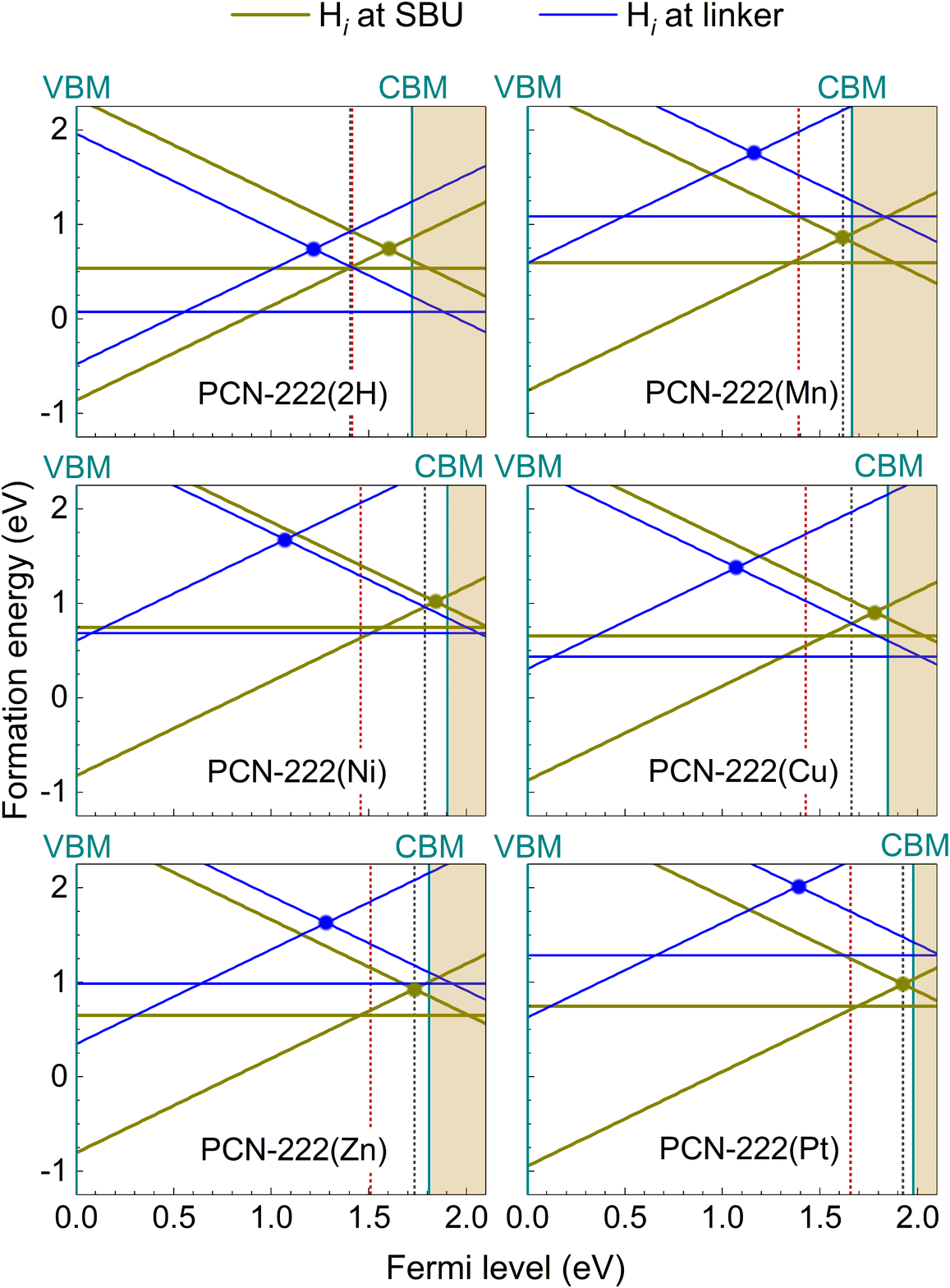}
\caption{Formation energies of H$_i$ in select MOFs. H$_i^+$ (H$_i^-$) have positive (negative) slopes and horizontal lines correspond to H$_i^0$. The local $\epsilon(+/-)$ levels of H$_i$ at the SBU and at the linker are marked by large solid red and blue dots, respectively; the global $(+/-)$ level (determined by the lowest-energy H$_i^+$ and H$_i^-$) and the effective $(+/-)$ level [i.e., the average of the local $\epsilon(+/-)$ levels at the SBU and the linker] are marked by the vertical gray and red dotted lines, respectively.} 
\label{fig;hi;fe} 
\end{figure}

Figure~\ref{fig;hi;fe} (and Figs.~S8 and S9) show the formation energy of H$_i$ in MOFs. In all the MOF compounds, H$_i$ defects show characteristics of  positive-$U$ defect centers with the charge-state transition level $\epsilon(+/0) < \epsilon(+/-) < \epsilon(0/-)$, except H$_i$ at the SBU in ZIF-8 which shows a weak negative-U character (see Fig.~S9). $U$ (not to be mistaken with $U$ in the PBE$+$$U$ method mentioned earlier) can be defined as the energy difference between $\epsilon(0/-)$ and $\epsilon(+/0)$ and is related to electron localization and structural relaxation.\cite{Anderson1975PRL} We find that $U$ of H$_i$ at the linker is larger than that at the SBU which indicates that at the linker the electronic states are more delocalized and/or the chemical bonding is more constrained (which results in a smaller relaxation energy when an electron is added to or removed from a charge state). Overall, the positive-$U$ character is consistent with the small difference in lattice relaxation observed between different charge states of H$_i$.   

In the non-zeolitic MOFs, H$_i^+$ is always energetically most favorable when the extra H$^+$ ion is bonded to an oxygen at the SBU, resulting in a protonated $\mu_2$-O or $\mu_3$-O unit, as seen in Figs.~\ref{fig;hi;struct}(a) and \ref{fig;hi;struct}(d)--(f). This is similar to H$_i^+$ in metal oxides where the proton is also attracted to oxygen.\cite{Peacock2003APL,Biswas2012PRL,Hoang2025PRM} H$_i^+$ also has the lowest formation energy in a wide range of Fermi-level values among all H$_i$ configurations at all possible lattice sites. The site preference of H$_i^0$ and H$_i^-$, on the other hand, is more sensitive to the specific metal species at the SBU and/or the linker. H$_i^0$ and H$_i^-$ are energetically most favorable at the linker in the case of PCN-222(2H) [see Fig.~\ref{fig;hi;struct}(b)], PCN-222($M$) [$M$ = Fe, Ni, or Cu; Fig.~\ref{fig;hi;struct}(c)], MOF-5 [Fig.~S7(d)], and UiO-66, and at the SBU in the case of PCN-222($M$) [$M$ = Mn, Zn, or Pt; similar to the H$_i^+$ configuration shown in Fig.~\ref{fig;hi;struct}(a)] and MIL-125 [similar to the H$_i^+$ configuration shown in Fig.~\ref{fig;hi;struct}(e)]. The reason for the difference can be ascribed to the electronic configuration of the metal ion at the linker in PCN-222: Mn$^{2+}$ has half-filled $d$ orbitals ($3d^5$), Zn$^{2+}$ is fully-filled ($3d^{10}$), and Pt$^{2+}$ has $5d$ orbitals which all make the addition of electron(s) to the linker a high-energy process. In the case of MIL-125, the ability of Ti$^{4+}$ to easily be reduced to Ti$^{3+}$ makes it a low-energy process to accommodate extra electrons in H$_i^0$ and H$_i^-$ at the Ti-based SBU. In zeolitic ZIF-8, H$_i^+$ and H$_i^-$ are energetically most favorable at the (nominal) SBU; Fig.~S9.

Defect levels induced by hydrogen interstitials in the band gap region of the MOF host include the {\it local} $\epsilon(+/0)$, $\epsilon(+/-)$, and $\epsilon(0/-)$ levels introduced by H$_i$ at the SBU and H$_i$ at the linker; see Fig.~\ref{fig;hi;fe} (and Figs.~S8 and S9). In the spirit of Van de Walle and Neugebauer's rationalization,\cite{VdW2003Nature} the local $\epsilon(+/-)$ level can be regarded as a probe of the chemical bonding at the SBU (or at the linker) and corresponding to the midpoint between the local anion-like and cation-like derived dangling bond states formed by the strong hydrogen defect--host interaction at the inorganic (organic) building block as described earlier. In addition, we define the {\it global} $(+/-)$ level as an energy level determined by the lowest-energy H$_i^+$ and H$_i^-$ configurations in the entire MOF. These H$_i^+$ and H$_i^-$ configurations are not necessarily the two charge states of the same defect; i.e., H$_i^+$ is at the SBU but H$_i^-$ can be at the SBU or the linker; see above. Note, however, that our purpose is not to investigate hydrogen interstitials as possible predominant point defects, but to use them as probes into the chemical bonding. In that context and in order to take into account the multi-component nature of the MOFs, we also introduce the {\it effective} $(+/-)$ level, obtained by taking the average of the {\it local} $\epsilon(+/-)$ levels at the SBU and at the linker. This effective hydrogen defect level is thus not a thermodynamic transition level.

Note that, in the materials under consideration, some local $\epsilon(0/-)$ levels are located above the CBM [see Figs.~\ref{fig;hi;fe} and S8], except those in ZIF-8 (Fig.~S9). In this case, the charge-state transition level occurs as a resonance in the conduction band and the negative charge state (H$_i^-$) may thus be thermodynamically unstable (although H$_i^-$ is structurally stable). As previously demonstrated and discussed in the literature,\cite{VdW2003Nature,Varley2024JAP} however, this should pose no problem for the formalism outlined by Van de Walle and Neugebauer\cite{VdW2003Nature} which our current approach is based upon.

\begin{figure}
\includegraphics[width=0.98\linewidth]{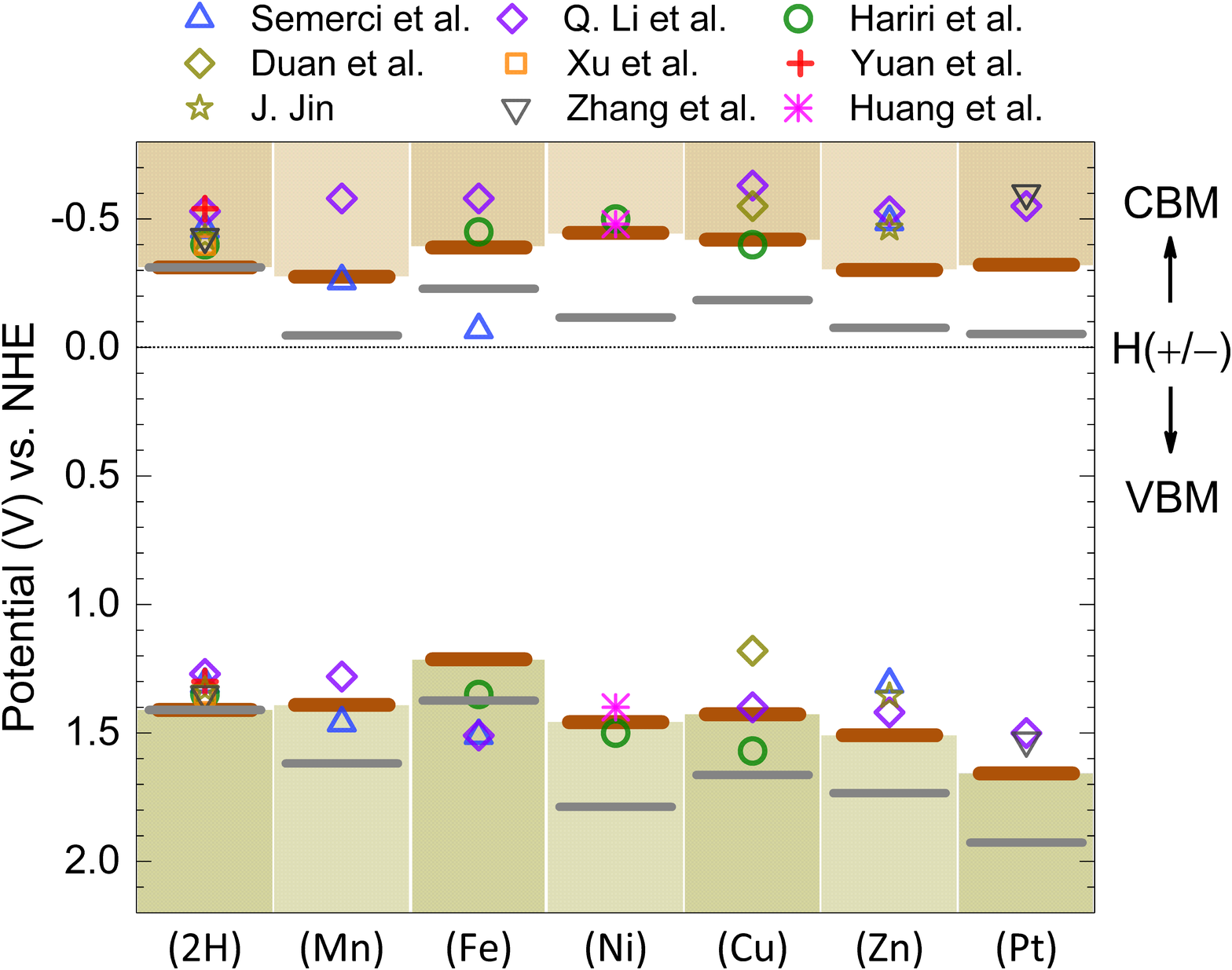}
\caption{Band edges of PCN-222(2H) and PCN-222($M$) based on the effective $(+/-)$ defect level (thick red bars) or the global $(+/-)$ defect level (thin gray bars), referred commonly to as the H$(+/-)$ level. Experimental band edges (symbols) are included for comparison.\cite{Semerci2023CT,Hariri2021AOC,Xu2015JACS,Jin2020NJC,Li2023P,Yuan2023CSA,Zhang2022JC,Duan2020APCS,Huang2024SCM}} 
\label{fig;align} 
\end{figure}

\begin{figure}
\includegraphics[width=0.98\linewidth]{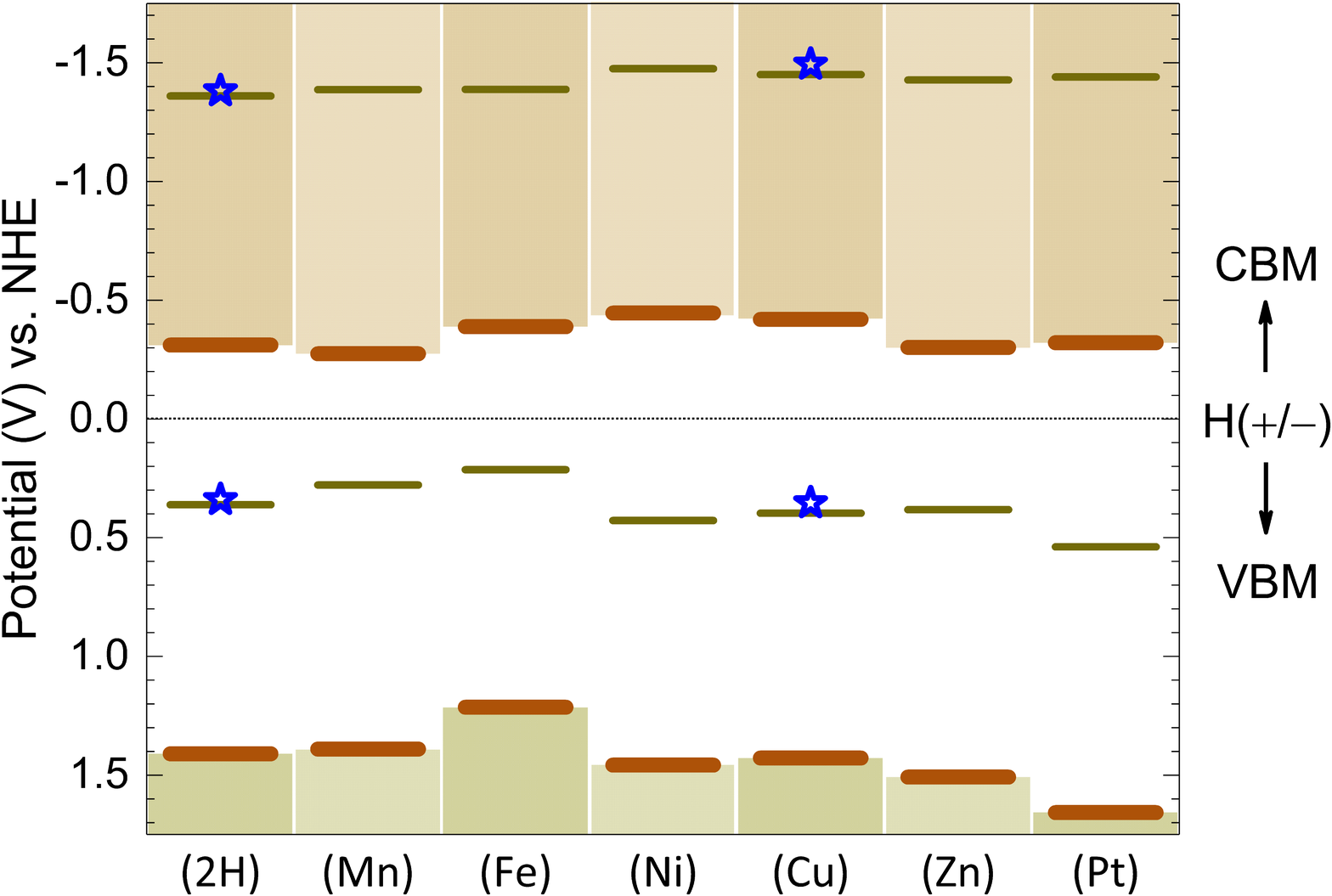}
\caption{Band edges of PCN-222(2H) and PCN-222($M$) obtained using different approaches involving (i) the effective $(+/-)$ defect level (thick red bars), (ii) bulk and slab calculations (symbols), and (iii) average electrostatic potentials at the center of the MOF pore (thin yellow bars).} 
\label{fig;align;compared} 
\end{figure}

\subsection{Electronic band alignment}\label{sec;results;align}

By assuming that the universal alignment of hydrogen levels\cite{VdW2003Nature} holds in the case of the MOFs, we set the global and effective $(+/-)$ levels [referred commonly to as the ``H$(+/-)$'' level] to the H$^+$(aq.)/H$_2$(g) electrode level. The calculated VBM and CBM can now be aligned on an absolute energy scale, e.g., with reference to the vacuum or the normal hydrogen electrode (NHE) level. Figure~\ref{fig;align} shows the band edges (vs.~NHE) of the PCN-222 series where $E_{\rm c }$ (V vs.~NHE) $=$ $-4.44$ $-$ $E_{\rm c}^{\rm {vac}}$ (eV vs.~vacuum); i.e., the global and effective $(+/-)$ levels are at 0 (V vs.~NHE). Our results, also listed in Table~S1, are in excellent agreement with experimentally reported band edges.\cite{Semerci2023CT,Hariri2021AOC,Xu2015JACS,Jin2020NJC,Li2023P,Yuan2023CSA,Zhang2022JC,Duan2020APCS,Huang2024SCM} More importantly, the results based on the effective $(+/-)$ level reproduce much better the overall chemical trend seen in experiments, compared to those obtained based on the global $(+/-)$ level. Note that the difference between the PCN-222 compounds is at the linker.

In experiments, the band-edge position of a semiconductor can be determined through the flatband potential ($V_{\rm {fb}}$) which is obtained in an electrochemical measurement where the material is in contact with an aqueous environment, i.e., involving a material/water interface. The measurement is performed at a pH value which corresponds to the point of zero charge (PZC), i.e., at which the interface has zero net charge. The calculated band edge (e.g., $E_{\rm c}$), on the other hand, involves a material/vacuum interface in the slab calculation approach, or the $(+/-)$ level in our current approach which is obtained in the absence of aqueous solutions in the MOF pores. The connection between $E_{\rm c}$ and $V_{\rm {fb}}$ for an $n$-type semiconductor can be described as\cite{Butler1978JES,Stevanovic2024PCCP}
\begin{equation}\label{eq;connection}
E_{\rm c} = V_{\rm {fb}} + \Delta_{\rm {fc}} + \Delta_{\rm {pH}} + \Delta_{\rm {dipole}},
\end{equation}
where $V_{\rm {fb}}$ is the flatband potential, i.e., the Fermi level of the material in the aqueous environment; $\Delta_{\rm {fc}}$ is the difference between the Fermi level and the CBM; $\Delta_{\rm {pH}}$ takes into account the pH dependence of the band-edge position in the environment, which can be approximated by the Nerst equation; and $\Delta_{\rm {dipole}}$ is related to interface dipoles that develop due to the interaction of the semiconductor with the aqueous solution. Among these terms, $\Delta_{\rm {pH}} = 0$ at the PZC, by definition. And since $V_{\rm {fb}} + \Delta_{\rm {fc}}$ is experimentally measured, the difference between the calculated and measured band edges is thus just $\Delta_{\rm {dipole}}$.

Returning to the theory-experiment comparison in Fig.~\ref{fig;align}, there we assume the measurements mentioned in the experimental reports were performed at the PZC; i.e., $\Delta_{\rm {pH}} = 0$. Also, the experimental data points are included as originally reported; i.e., no adjustment even when some authors assumed the CBM was at the flatband $V_{\rm {fb}}$, i.e., $\Delta_{\rm {fc}} = 0$, and others assumed $V_{\rm {fb}}$ was below the CBM by 0.1 or 0.2 eV.\cite{Semerci2023CT,Hariri2021AOC,Xu2015JACS,Jin2020NJC,Li2023P,Yuan2023CSA,Zhang2022JC,Duan2020APCS,Huang2024SCM} Finally, the effect of the interface dipoles is not included in our calculation. It can, however, be estimated by using a well known case in experiments. For example, we can align the calculated CBM of PCN-222(2H) to the experimental values, which gives $\Delta_{\rm {dipole}}$ a value of $0.1$--$0.2$ eV. Note that the error bar of our calculations is about 0.1 eV. Also, $E_{\rm c}$ is compared directly with that obtained in flatband potential measurements. Any discrepancies among the measured band gaps (often determined from absorption spectroscopy) and between the experimental data and the calculated values would affect only $E_{\rm v}$.  

\begin{figure}
\includegraphics[width=0.98\linewidth]{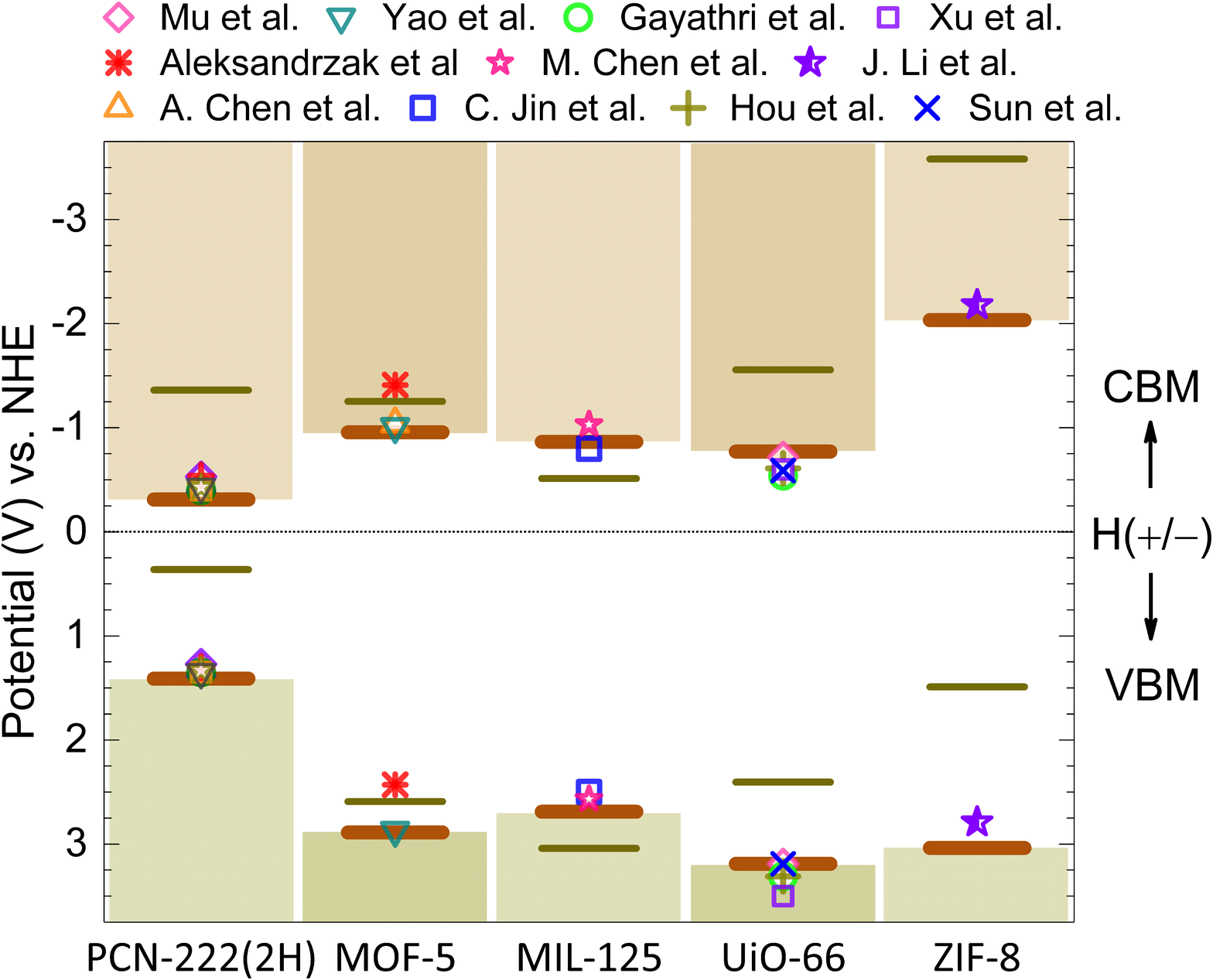}
\caption{Band edges of MOF-5, MIL-125, UiO-66, and ZIF-8 based on the effective $(+/-)$ defect level (thick red bars) and the average pore potential approach (thin yellow bars), and reported experimental values (symbols).\cite{Aleksandrzak2020RSCA,Chen2021RKMC,Yao2022JSSC,Chen2020IECR,Jin2024RSCAdv,Dai2025MD,Hou2022CST,Sun2024IC,Mu2018DT,Gayathri2022CSA,Xu2018ACSC,Li2022CST} The data for PCN-222(2H) is repeated for comparison.}
\label{fig;various} 
\end{figure} 

Figure~\ref{fig;align;compared} compares the calculated band edges of the PCN-222 series obtained using three different computational approaches. The approaches based on bulk and surface calculations and on average electrostatic potentials at the MOF pore center provide almost the same results in the case of PCN-222(2H) and PCN-222(Cu). These two approaches, however, both give the band-edge values much higher on an absolute energy scale than those based on the universal alignment of the effective $(+/-)$ level (by 1.00--1.13 eV for PCN-222) and than the experimental values.

Finally, to check the robustness of our approach, we apply the same procedure to MOF-5, MIL-125, UiO-66, and ZIF-8, which are significantly different structurally and chemically from the PCN-222 series. ZIF-8 is selected in particular for its reportedly very high conduction-band edge on an absolute energy scale, compared to other MOFs.\cite{Butler2014JACS} Figure~\ref{fig;various} shows excellent agreement between the results obtained based on the effective $(+/-)$ level, also listed in Table~S2, and those from experimentally measured band edges. The average pore-center potential approach appears to work fine for simple MOFs such as MOF-5 and MIL-125 but gives band edges that are about 1.55 eV (0.79 eV) higher than the experimental values in the case of ZIF-8 (UiO-66). As is evident in Fig.~\ref{fig;various}, the discrepancy cannot be fixed by a general shift. These results, again, show that the effective $(+/-)$ level of the hydrogen interstitials can be used as a CNL and our approach performs consistently across MOFs with different SBUs and linkers.

It should be noted that Van de Walle and Neugebauer's original work on universal alignment focuses primarily on tetrahedrally coordinated semiconductors without electron lone pairs.\cite{VdW2003Nature} In materials with anion lone pairs discussed in the literature, such as $\beta$-Ga$_2$O$_3$ and SnO$_2$ with three-coordinated oxygen atoms where H$_i^+$ is strongly bonded to a lone pair at the $\mu_3$-like oxygen but ``without much influence on the lattice,''\cite{Swallow2019APLM,Varley2010APL} the lone-pair H$_i^+$ configuration may not actually probe the anion- and cation-derived dangling bond states. As a result, a $(+/-)$ level associated with such an H$_i^+$ configuration may not correspond to a CNL, and one may expect an offset between $(+/-)$ levels in materials with and without anion lone pairs.\cite{Swallow2019APLM} In the MOFs, however, H$_i^+$ at the $\mu_2$- or $\mu_3$-O atom strongly disturbs the local anion-like--cation-like bonds as described earlier, unlike in $\beta$-Ga$_2$O$_3$. Further studies might still be needed to quantify any offset, if exists, between the $(+/-)$ levels in the MOFs and in tetrahedrally coordinated materials.

\section{Conclusion}

We have developed a general approach to align the electronic band structure of MOFs on an absolute energy scale which can be used for direct comparison with band edges measured in experiments. In this approach, hydrogen interstitial defects are employed as probes into the chemical bonding of the host materials. The effective $(+/-)$ energy level, defined as the average of the $\epsilon(+/-)$ charge-state transition levels of hydrogen interstitials at the SBU and at the linker of a MOF, is identified as a charge neutrality level which can be used to line up the band edges. This effective hydrogen defect level captures the subtle chemical details of both the building blocks in the hybrid inorganic-organic materials. The results based on this approach for a series of similar as well as widely different MOFs show excellent agreement with reported electrochemically measured band-edge positions. Our approach is conceptually general and applicable to not just MOFs but can also be extended to other complex, hybrid/multi-component materials. It will facilitate efficient screening and design of complex materials with proper electronic band edges for photocatalytic, photovoltaic, and electrochemical applications.  

\begin{acknowledgments}

The author is grateful to Chris Van de Walle for insightful discussions. This work used resources of the Center for Computationally Assisted Science and Technology (CCAST) at North Dakota State University, which were made possible in part by U.S. National Science Foundation MRI Award No.~2019077.

\end{acknowledgments}

\appendix

\section{Supplementary Information}\label{sec;app}

\renewcommand{\thetable}{S1}
\begin{table*}
\caption{Structural and electronic properties of PCN-222(2H) and PCN-222($M$). }\label{tab;bulk}
\begin{center}
\begin{ruledtabular}
\begin{tabular}{lrrrrrrr}
&(2H) &(Mn) &(Fe) &(Ni) &(Cu) &(Zn) &(Pt) \\
\hline
PBE$+$$U$ (eV) & 0 & 10.0 & 6.0 & 6.0 & 7.0 & 0 & 6.0 \\
$M$ at linker & & Mn$^{2+}$ $3d^5$ & Fe$^{2+}$ $3d^6$ & Ni$^{2+}$ $3d^8$ & Cu$^{2+}$ $3d^9$ & Zn$^{2+}$ $3d^{10}$ & Pt$^{2+}$ $5d^{8}$   \\
Magn.~mo.~($\mu_{\rm B}$) && 4.85 & 3.81 & 0 & 0.68 & 0 & 0\\
$a$, $b$ ({\AA}) &43.40 &43.43 &43.39 &43.31 &43.38 &43.37 & 43.38 \\
$c$ ({\AA}) &16.90 &16.92 &16.92 &16.90 &16.85 &16.92 & 16.92 \\
$M$--N (\AA) && 2.09, 2.11 & 2.06, 2.07 & 1.98, 2.00 & 2.01, 2.03 & 2.05, 2.06 & 2.02, 2.03 \\
$E_g$ (eV) & 1.72 & 1.67 & 1.60 & 1.90 & 1.85 & 1.81 & 1.98 \\
$E_g$, expt (eV) & 1.75--1.84$^a$ & 1.72--1.86$^b$ & 1.58--2.09$^c$ & 1.88--2.00$^d$ & 1.73--2.03$^e$ & 1.80--1.95$^f$ & 2.05--2.13$^g$ \\
$E_{\rm v}$ (V vs.~NHE) & 1.41 & 1.39 & 1.21 & 1.46 & 1.43 & 1.51 & 1.66 \\  
$E_{\rm c}$ (V vs.~NHE) & $-$0.31 & $-$0.28 & $-$0.39 & $-$0.45 & $-$0.42 & $-$0.30 & $-$0.32 \\
\end{tabular}%
\end{ruledtabular}
\end{center}
\begin{flushleft}
$^a$Refs.~\cite{Semerci2023CT,Li2023P,Hariri2021AOC,Xu2015JACS,Yuan2023CSA,Jin2020NJC,Zhang2022JC}. $^b$Refs.~\cite{Semerci2023CT,Li2023P}. $^c$Refs.~\cite{Semerci2023CT,Li2023P,Hariri2021AOC}. $^d$Refs.~\cite{Hariri2021AOC,Huang2024SCM}. $^e$Refs.~\cite{Li2023P,Hariri2021AOC,Duan2020APCS}. $^f$Refs.~\cite{Semerci2023CT,Li2023P,Jin2020NJC}. $^g$Refs.~\cite{Li2023P,Zhang2022JC}.
\end{flushleft}
\end{table*}

\renewcommand{\thetable}{S2}
\begin{table}
\caption{Structural and electronic properties of MOF-5, MIL-125, UiO-66, and ZIF-8.}\label{tab;bulk2}
\begin{center}
\begin{ruledtabular}
\begin{tabular}{lrrrr}
&MOF-5 &MIL-125 &UiO-66 &ZIF-8 \\
\hline
HSE($\alpha$) & 0.05 & 0.22 & 0.18 & 0.20 \\
$a$, $b$ ({\AA}) & 26.12 & 19.24 & 20.94 & 17.04 \\
$c$ ({\AA}) & & 18.21 & &  \\
$E_g$ (eV) & 3.84 & 3.55 & 3.96 & 5.07  \\
$E_g$, expt (eV) & 3.84--3.88$^a$ & 3.30--3.60$^b$ & 3.78--4.10$^c$ & 4.97$^d$  \\
$E_{\rm v}$ (V, vs.~NHE) & 2.89 & 2.69 & 3.19 & 3.04 \\  
$E_{\rm c}$ (V, vs.~NHE) & $-$0.96 & $-$0.86 & $-$0.77 & $-$2.04  \\
\end{tabular}%
\end{ruledtabular}
\end{center}
\begin{flushleft}
$^a$Refs.~\cite{Yao2022JSSC,Aleksandrzak2020RSCA}. $^b$Refs.~\cite{Jin2024RSCAdv,Chen2020IECR}. $^c$Refs.~\cite{Mu2018DT,Gayathri2022CSA,Hou2022CST,Xu2018ACSC,Sun2024IC}. $^d$Ref.~\cite{Li2022CST}.
\end{flushleft}
\end{table}

\renewcommand{\thefigure}{S1}
\begin{figure*}
\centering
\includegraphics[width=0.8\linewidth]{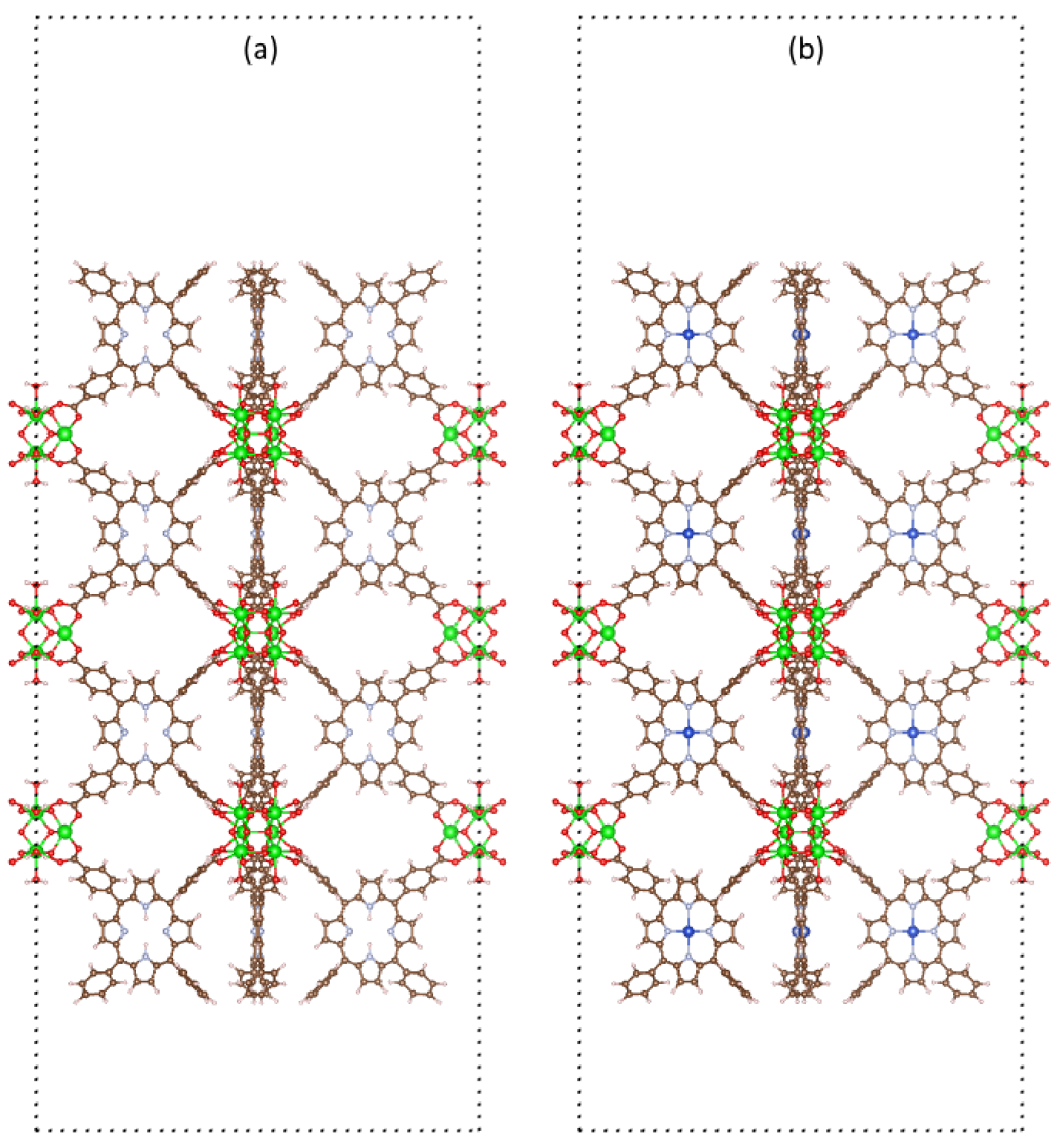}
\caption{Slab models for the (001) surface of PCN-222(2H) and PCN-222(Cu). Color code: red = O, brown = C, pink = H, light blue = N, green = Zr, and blue = Cu.}
\label{fig;slabs}
\end{figure*}

\renewcommand{\thefigure}{S2}
\begin{figure*}
\vspace{2.0cm}
\centering
\includegraphics[width=0.95\linewidth]{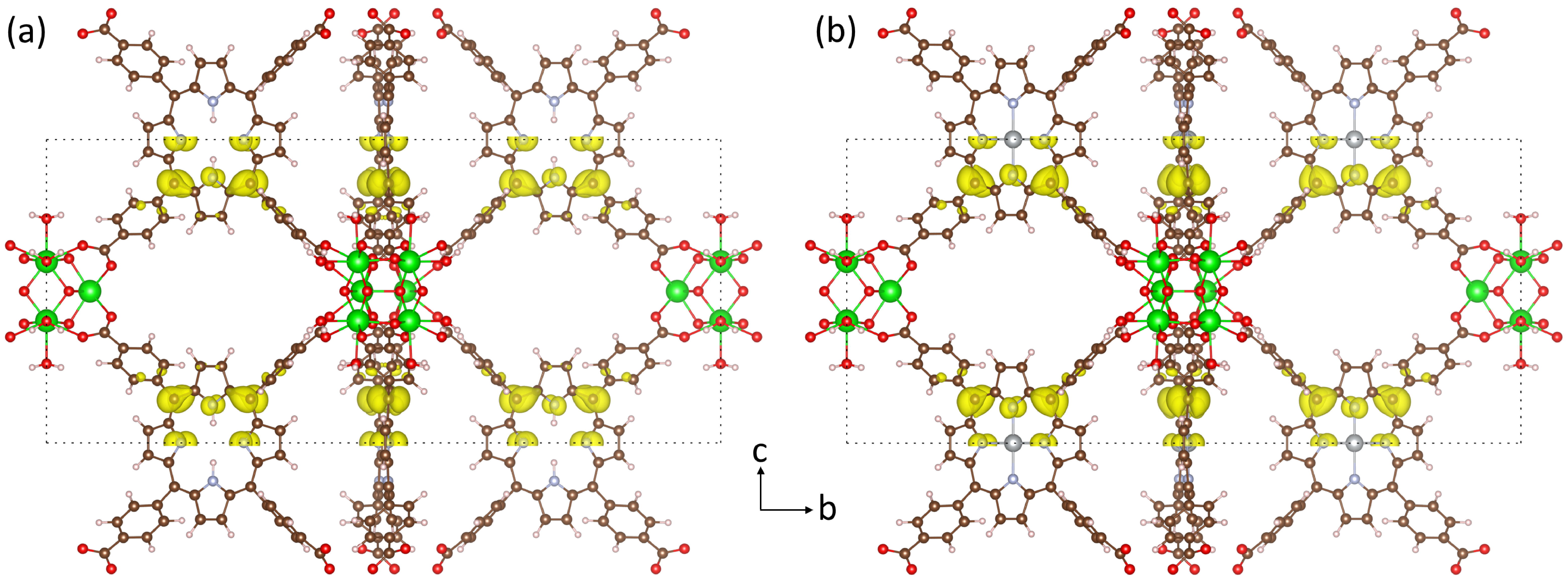}
\caption{Structures of (a) PCN-222(2H) and (b) PCN-222(Ni) and charge densities associated with the highest occupied state (VBM). Color code: red = O, brown = C, pink = H, light blue = N, green = Zr, and gray = Ni.}
\label{fig;struct} 
\end{figure*} 


\renewcommand{\thefigure}{S3}
\begin{figure*}
\vspace{3cm}
\centering
\includegraphics[width=0.95\linewidth]{figS3}
\caption{Total and atom-decomposed electronic densities of states near the band-gap region in PCN-222($M$), $M$ = Pt, Mn, Fe, Ni, Cu, and Zn. The zero of energy is set to the highest occupied state.} 
\label{fig;dosx} 
\end{figure*}


\renewcommand{\thefigure}{S4}
\begin{figure*}
\vspace{5cm}
\centering
\includegraphics[width=0.95\linewidth]{figS4}
\caption{Total and atom-decomposed electronic densities of states near the band-gap region in MOF-5, MIL-125, UiO-66, and ZIF-8. The zero of energy is set to the highest occupied state.} 
\label{fig;dos2} 
\end{figure*}


\renewcommand{\thefigure}{S5}
\begin{figure*}
\vspace{4.5cm}
\centering
\includegraphics[width=0.90\linewidth]{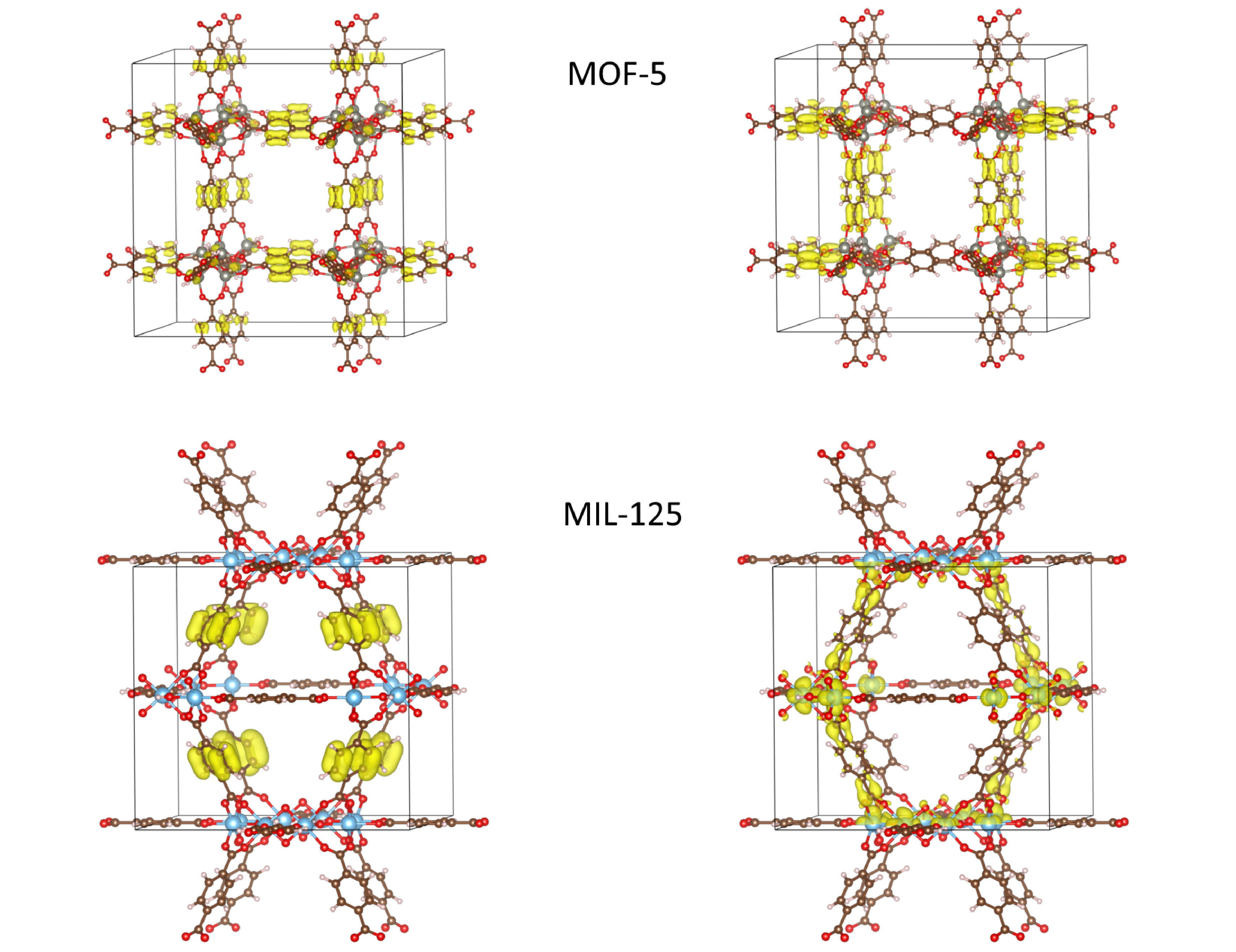}
\caption{Structures of MOF-5 and MIL-125 and charge densities associated with the highest occupied state (VBM; left) and the lowest unoccupied state (CBM; right). Color code: red = O, brown = C, pink = H, gray = Zn, and blue = Ti.}
\label{fig;struct1}
\end{figure*}

\renewcommand{\thefigure}{S6}
\begin{figure*}
\vspace{4.5cm}
\centering
\includegraphics[width=0.9\linewidth]{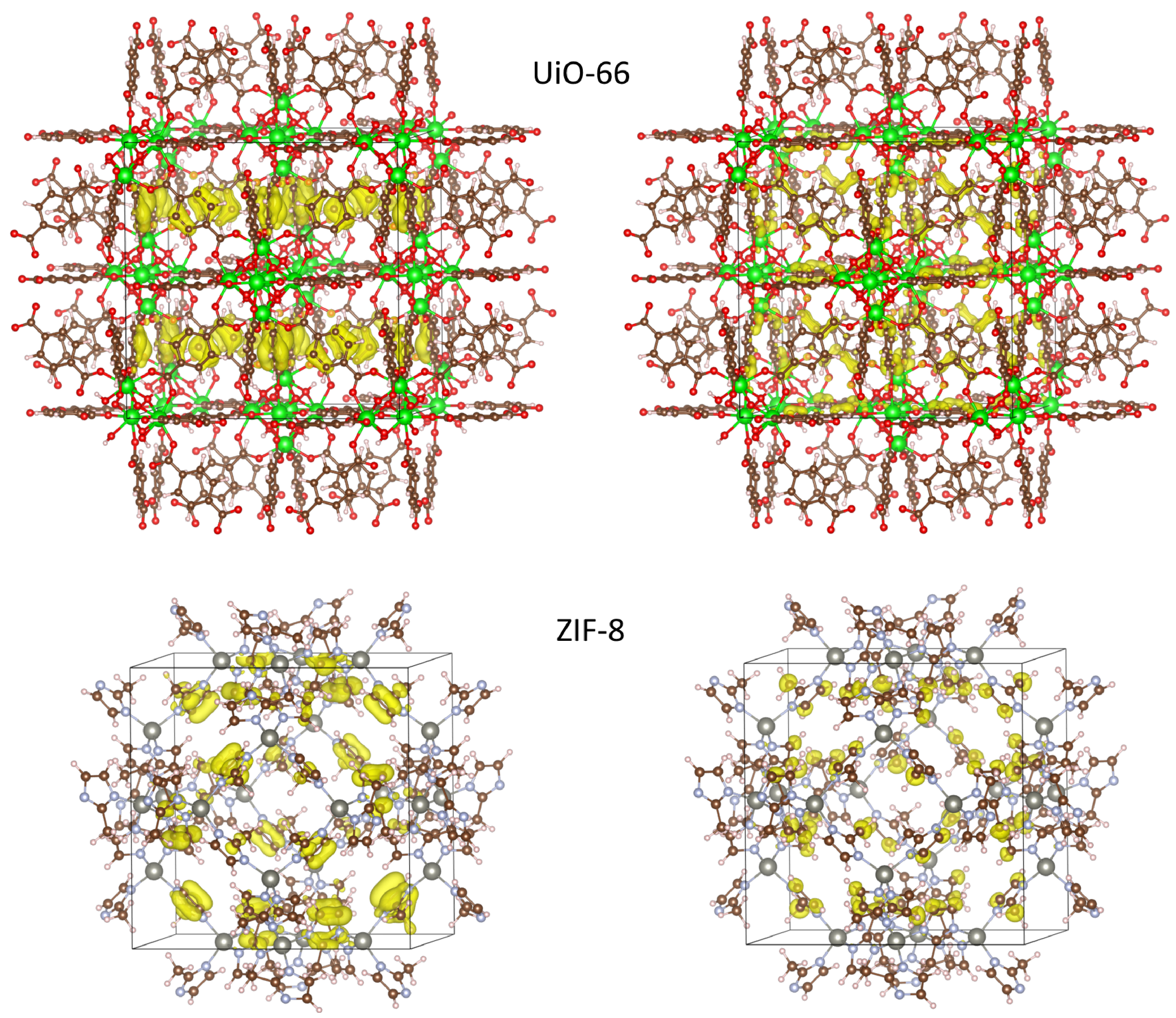}
\caption{Structures of UiO-66 and ZIF-8 and charge densities associated with the highest occupied state (VBM; left) and the lowest unoccupied state (CBM; right). Color code: red = O, brown = C, pink = H, green = Zr, and gray = Zn.}
\label{fig;struct2}
\end{figure*}


\renewcommand{\thefigure}{S7}
\begin{figure*}
\vspace{3.8cm}
\centering
\includegraphics[width=0.9\linewidth]{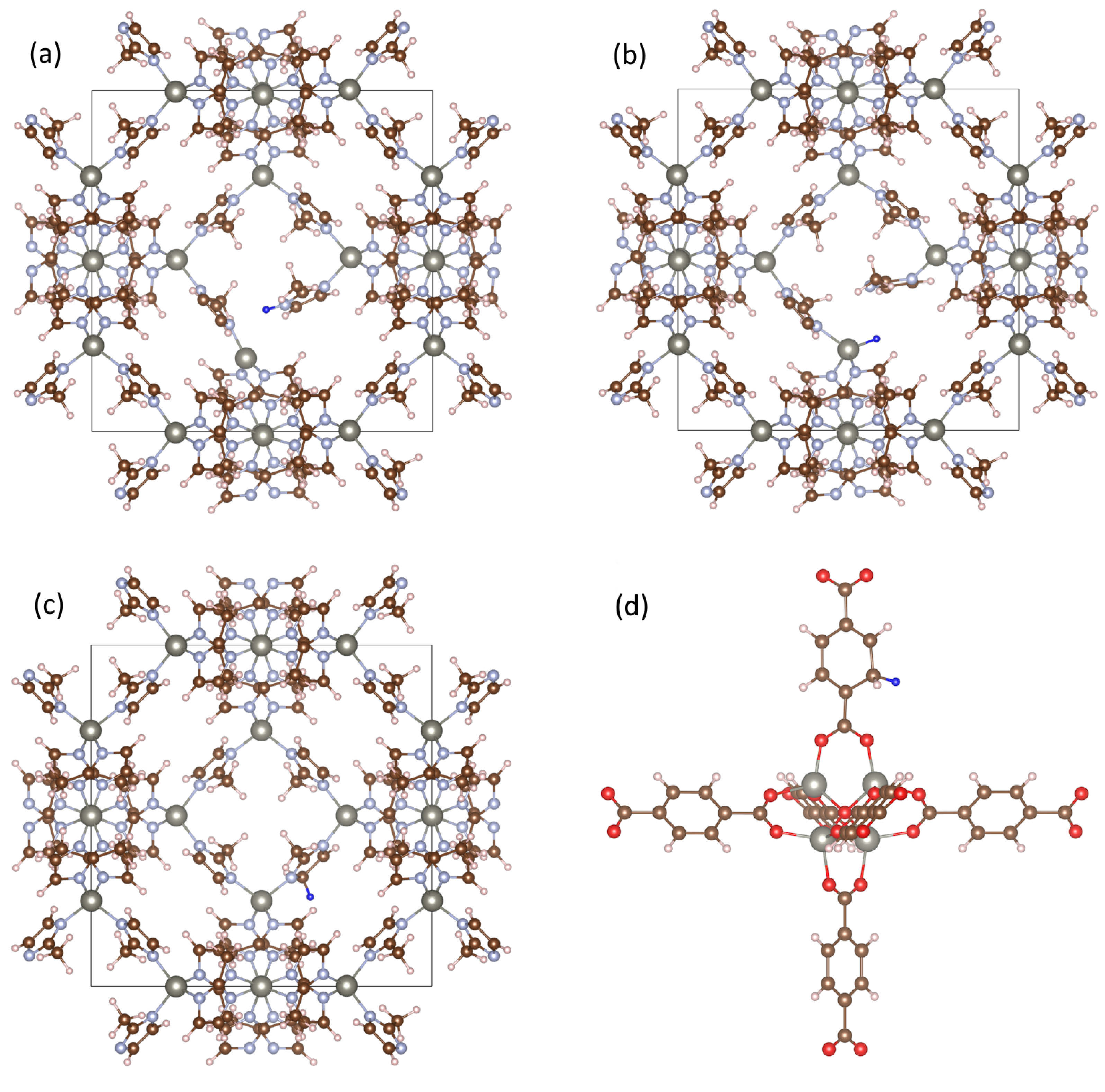}
\caption{Local structure of (a) H$_i^+$ and (b) H$_i^-$ at the SBU and (c) H$_i^0$ at the 2-mIM linker in ZIF-8, and (d) H$_i^-$ at the BDC linker in MOF-5. H$_i$ is represented by a small blue sphere. Other H$_i$ defects at the BDC linker in MOF-5, MIL-125, and UiO-66 have a similar structure to H$_i^-$ in MOF-5. Color code: red = O, brown = C, pink = H, light blue = N, and gray = Zn.}
\label{fig;nh2}
\end{figure*}

\renewcommand{\thefigure}{S8}
\begin{figure*}
\includegraphics[width=0.55\linewidth]{figS8}
\caption{Formation energies of hydrogen interstitials (H$_i$), as a function of the Fermi level, in PCN-222(Fe), MOF-5, MIL-125, and UiO-66. H$_i^+$ (H$_i^-$) have positive (negative) slopes and horizontal lines correspond to H$_i^0$. The local $\epsilon(+/-)$ levels of H$_i$ at the SBU and at the linker are marked by large dark yellow and blue solid dots, respectively; the global $(+/-)$ level (determined by the lowest-energy H$_i^+$ and H$_i^-$) and the effective $(+/-)$ level (i.e., the average of the local $\epsilon(+/-)$ levels at the SBU and the linker) are marked by the vertical black and red dotted lines, respectively. In this presentation, the hydrogen chemical potential $\mu_{\rm H}$ is set to $-0.23$ eV [in the case of MOF-5 and UiO-66] or $-1.60$ eV [PCN-222(Fe) and MIL-125].} 
\label{fig;fe} 
\end{figure*}

\renewcommand{\thefigure}{S9}
\begin{figure*}
\includegraphics[width=0.55\linewidth]{figS9}
\caption{Formation energies of hydrogen interstitials (H$_i$) in ZIF-8. The local $\epsilon(+/-)$ levels of H$_i$ at the SBU and at the linker are marked by large solid dark yellow and blue dots, respectively; the global (effective) $(+/-)$ level is marked by the vertical black (red) dotted line. In this presentation, the hydrogen chemical potential $\mu_{\rm H}$ is set to $-0.23$ eV; see text.} 
\label{fig;fe;zif} 
\end{figure*}


%

\end{document}